\documentclass[
amsmath,amssymb,twocolumn,aps,
]{revtex4-2}
\usepackage{graphicx,graphics,color,epsfig}
\usepackage{bm}
\usepackage{dcolumn}
\usepackage{amsmath}
\usepackage{amssymb}
\usepackage{graphicx}
\usepackage{epstopdf}
\usepackage[colorlinks,
anchorcolor=blue,  
linkcolor=blue,       
citecolor=blue]{hyperref}  
\usepackage{natbib}
\usepackage{multirow}  
\usepackage{lineno} 
\usepackage{ulem}

\newcommand{\be}{\begin{eqnarray}}
\newcommand{\ee}{\end{eqnarray}}

\newcommand{\ba}{\begin{align}}
\newcommand{\ea}{\end{align}}

\makeatletter

\newcommand{\Rmnum}[1]{\expandafter\@slowromancap\romannumeral #1@}
\makeatother

\begin{document}
\title{Varying quench dynamics in the transverse Ising chain:\\ the Kibble-Zurek, saturated and pre-saturated regimes}
\author{Han-Chuan Kou}
\affiliation{College of Physics, Sichuan University, 610064, Chengdu, People’s Republic of China\\and Key Laboratory of High Energy Density Physics and Technology of Ministry of Education, Sichuan University, 610064,Chengdu, People’s Republic of China}
\author{Peng Li}
\email{lipeng@scu.edu.cn}
\affiliation{College of Physics, Sichuan University, 610064, Chengdu, People’s Republic of China\\and Key Laboratory of High Energy Density Physics and Technology of Ministry of Education, Sichuan University, 610064,Chengdu, People’s Republic of China}

\date{\today}

\begin{abstract}
   According to the Kibble-Zurek mechanism, there is a universal power-law relationship between the defect density and the quench rate during a slow linear quench through a critical point. It is generally accepted that a fast quench results in a deviation from the Kibble-Zurek scaling law and leads to the formation of a saturated plateau in the defect density. By adjusting the quench rate from slow to very fast limits, we observe the varying quench dynamics and identify a pre-saturated regime that lies between the saturated and Kibble-Zurek regimes. This significant result is elucidated through the adiabatic-impulse approximation first, then verified by a rigorous analysis on the transverse Ising chain as well. As we approach the turning point from the saturated to pre-saturated regimes, we notice a change in scaling laws and, with an increase in the initial transverse field, a shrinking of the saturated regime until it disappears. During another turning point from the Kibble-Zurek to pre-saturated regimes, we observe an attenuation of the dephasing effect and a change in the behavior of the kink-kink correlation function from a Gaussian decay to an exponential decay. Finally, the coherent many-body oscillation after quench exhibits different behaviors in the three regimes and shows a significant change of scaling behavior between the S and PS regimes.
\end{abstract}

\maketitle


\section{Introduction}

The Kibble-Zurek mechanism (KZM) describes how topological defects form in a system undergoing a continuous phase transition at a finite rate  \cite{Kibble_1976, Kibble_1980, Zurek_1985, Zurek_1993, Zurek_1996}. It has been widely applied in condensed matter physics, becoming one of the cornerstones of non-equilibrium dynamics and leading to numerous experimental tests \cite{Isaac_1991, Mark_1994, Ruutu_1996, Pickett_1996, Monaco_2002, Maniv_2003, Sadler_2006, Weiler_2008, Golubchik_2010, Chiara_2010, Griffin_2012, Chomaz_2015, Yukalov_2015, Zoran_2015}. In recent years, the quantum KZM (QKZM), a quantum version of KZM, has attracted significant interest for its application to quenches across a quantum critical point \cite{Damski_2005, Zurek_2005, Polkovnikov_2005, Dziarmaga_2005, Dziarmaga_2010}. The QKZM predicts that the defect density scales as $n\propto\tau_{Q}^{-d\nu/(1+z\nu)}$ in terms of the equilibrium critical exponents, where $\tau_Q$ is the quench time, $d$ is the dimensionality of the system, and $z$ and $\nu$ are dynamical exponent and correlation length exponent respectively. This scaling law holds for slow quench and the quench time sets the KZ length scale. Both theoretical \cite{Levitov_2006, Uwe_2006, Zurek_2007-2, Zurek_2007, Hiroki_2007, Sen_2007, Sen_2008, Sen_2008-2, Sen_2008-3, Polkovnikov_2008, Zurek_2008, Dutta_2009, Zurek_2010, Zurek_2013, Sen_2014, Dutta_2017, Dziarmaga_2019, Zurek_2019, Zurek_2020, Divakaran_2020, Vicari_2020, Kormos_2020, Damski_2020, Dziarmaga_2021, Kou_2022, Sim_2022, Dziarmaga_2022, Dziarmaga_2023} and experimental \cite{Chen_2011, Baumann_2011, Singer_2013, Guo_2014, Braun_2015, Anquez_2016, Carolyn_2016, Guo_2016, Cheng_2016, Zurek_2018, Sachdev_2019, Suzuki_2020, Weinberg_2020, Guo_2020} research in this area has made tremendous progress.

It is now widely accepted that fast quenches will eventually result in deviations from the KZM predictions. For instance, saturated plateaus instead of the KZ scaling law in the defect density have been uncovered in confined ion chains \cite{Campo_2010, Chiara_2010}, the holographic superconducting ring \cite{Zurek_2015}, and the one-dimensional quantum ferromagnet \cite{Campo_2019}. The breakdown of KZ scaling law stimulates subsequent theoretical investigations \cite{Liu_2015, Sun_2021, Campo_2023, Campo_2023-2, Campo_2023-3}. The appearance of plateaus in the defect density has also been confirmed by experimental evidences in the ultracold Bose atoms and Fermi gases, in which the systems are driven through the quantum phase transition at a fast or moderate quench rate \cite{Dziarmaga_2013, Donadello_2016, Proukakis_2018, Shin_2019, Shin_2021, Shin_2022, shin_2023}. An empirical formula is conjectured to fit the experimental data near the change from KZ scaling to the saturated plateau \cite{Donadello_2016, Shin_2019, shin_2023}. Subsequently, various studies demonstrate that the occurrence of the plateau can be ascribed to the early-time coarsening before the freeze-out time and the universality in the deviation from KZM is established \cite{Campo_2023, Campo_2023-2, Campo_2023-3}.

From the point of view of the sudden quench, it is natural to envisage the appearance of a saturated regime since there is an upper bound for the defect density \cite{Sen_2008-4, Sarkar_2020, Mainak_2020}. However, there is a lack of quantitative studies on the detailed variation of quench dynamics. A few works showed there may be an intermediate regime between the KZ and the saturated regimes. In a study of holographic superfluids, it was shown that the fast and very fast quenches can lead to distinguishable behaviors based on the comparison of the final time, freeze-out time, and the timescale in which the order parameter grows \cite{Liu_2015}. In another study within the framework of conformal field theory, the authors established new scaling behaviors that may dominate the intermediate regime \cite{Das_2016, Das_2017-2, Sun_2021}.

In this work, we focus on the density of kinks and the kink-kink correlation function in the one-dimensional transverse Ising chain, which have been recently studied experimentally \cite{King_2022}. Notably, we provide conclusive evidence for the existence of an intermediate regime between the Kibble-Zurek (KZ) and saturated (S) regimes through this prototypical model. Here, the intermediate regime is referred to as the pre-saturated (PS) regime, since it shares some common features with the saturated one. We establish precise formula of defect density in the PS regime, which goes beyond the empirical one in Ref. \cite{Donadello_2016, Shin_2019, shin_2023}.  There are two turning points. One labels the breakdown of the S scaling law from S to PS regimes, where we observe a change in scaling laws and a shrinking of the saturated regime until it disappears with the initial transverse field increasing. Another one labels the breakdown of the KZ scaling law from KZ to PS to regimes, where we observe an attenuation of the dephasing effect and a change in the behavior of the kink-kink correlation function from a Gaussian decay to an exponential decay.

The paper is organized as follows. In Sec. \ref{Sec-AI}, we show the scenario of adiabatic-impulse (AI) approximation from  slow to fast quenches, which tells us briefly why there can be a PS regime between the KZ and S regimes. In Sec. \ref{Subsec-Ising}, the linear quench protocol for the transverse Ising chain is established. In Sec. \ref{Subsec-density}, we elaborate on the quench dynamics in the three regimes, which shows clear both analytical and numerical evidences for the existence of the PS regime. The two turning points therein are also discussed in detail. In Secs. \ref{Sec-kkcor} and \ref{Sec-coherent}, we study kink-kink correlation function and many-body oscillation respectively. At last, we give a summary in Sec. \ref{Sec-discussion}.

\section{Adiabatic-impulse approximation}\label{Sec-AI}

Firstly, we start from the AI approximation, which is applicable to a variety of systems with second-order phase transition \cite{Dziarmaga_2010}. A system is linearly ramped from $g(t_i)=g_i$ to $g(t_f)=g_f(<g_i)$ across a critical point $g_c$ at a rate characterized by a quench time $\tau_Q$, where $g(t)=-\frac{t}{\tau_Q}$ is the parameter of the system, and $t_i$ and $t_f$ are the initial and final times. A distance from a quantum critical point can be measured with a dimensionless parameter $\epsilon(t)=\frac{g(t)-g_c}{g_c}$.

\begin{figure}[t]
  \begin{center}
		\includegraphics[width=3.22in,angle=0]{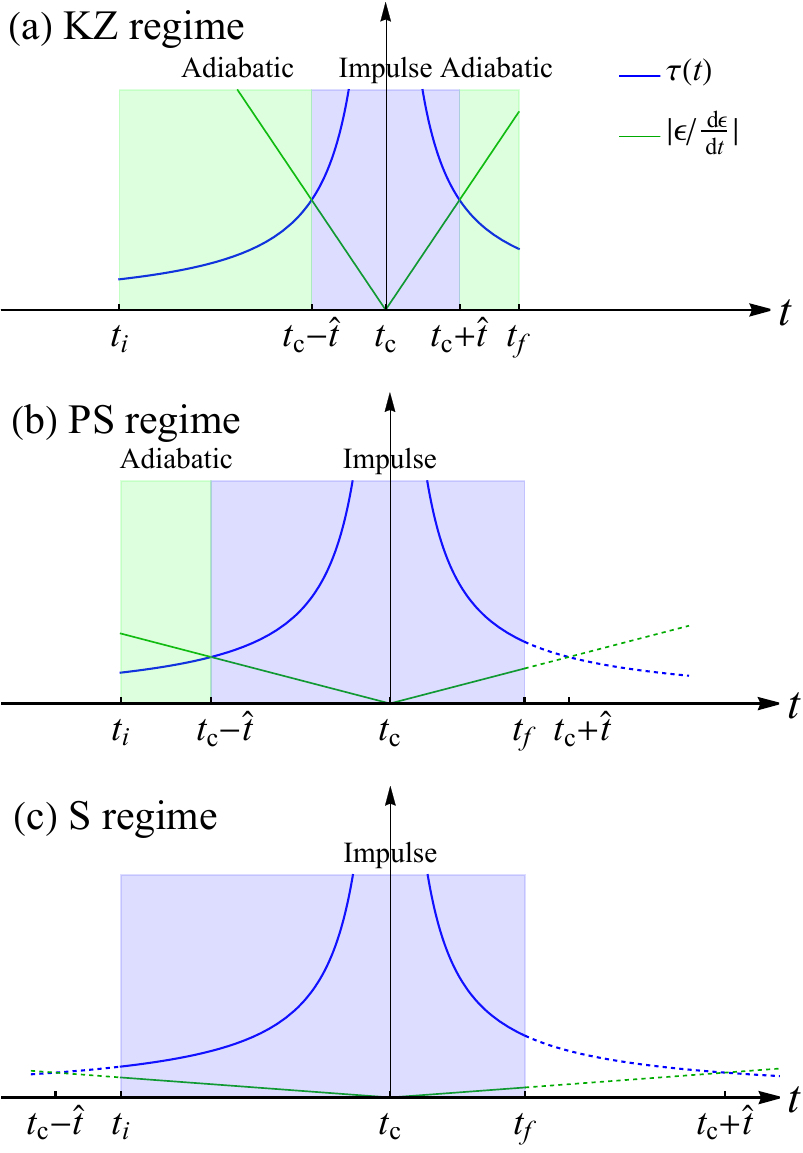}
  \end{center}
  \caption{The scenario of AI approximation: KZ regime (a), PS regime (b), and S regime (c). The blue line represents the relaxation time $\tau(t)$ and the green line represents the inverse the transition rate of linear quench $\left|\epsilon/\frac{d\epsilon}{dt}\right|$. In (b) and (c), the extended dash lines are used to determine the intersection points between $\tau(t)$ and $\left|\epsilon/\frac{d\epsilon}{dt}\right|$ outside of linear quench (i.e., $t<t_i$ or $t>t_f$).}
	\label{plot-AIapprox}
\end{figure}

Generally speaking, the system can be prepared far away from the critical point to acquire a simple initial state. After quench, the system is driven to a final state, in which the defects due to critical dynamics are easy to be counted. So it is reasonable to assume
\begin{equation}
  |g_i-g_c|\gg |g_f- g_c|. \label{assumption}
\end{equation}
The system evolves non-adiabatically in the time interval $t_c-\hat{t}<t<t_c+\hat{t}$, where $t_c=-g_c\tau_Q$ is the time when the system crosses the critical point, $\hat{t}\propto\tau_Q^{z\nu/(1+z\nu)}$ the frozen-out time, $z$ the dynamical exponents, and $\nu$ correlation length exponents. The frozen-out time is a special time scale at which the relaxation time $\tau(\hat{t})$ equals the inverse transition rate of linear quench $\left|\epsilon/\frac{d\epsilon}{dt}\right|_{t=\hat{t}}$ where $\epsilon=\frac{g(t)-g_c}{g_c}$ is the distance from the critical point.  Here, we can write the three time scales: the initial time $t_i=-g_i\tau_Q$, the final time $t_f=-g_f\tau_Q$, and the frozen-out time $\hat{t}$. In the KZ regime, we get the time scales approximately sequenced as
\begin{equation}
  t_i<t_c-\hat{t}<t_c+\hat{t}<t_f. \label{KZreg}
\end{equation}
As $\tau_Q$ decreases, the time scales becomes sequenced as
\begin{equation}
  t_i<t_c-\hat{t}<t_f<t_c+\hat{t}, \label{PSreg}
\end{equation}
which corresponds to the PS regime. As $\tau_Q$ decreases further, the time scales becomes sequenced as
\begin{equation}
  t_c-\hat{t}<t_i<t_f<t_c+\hat{t}, \label{Sreg}
\end{equation}
where the system enters into the S regime. The full scenario is illustrated in Fig. \ref{plot-AIapprox}. Moreover, the turning points can also be estimated in the framework of  the AI approximation. First, by setting $t_c+\hat{t}=t_f$, we can estimate a quench time scale
\begin{equation}\label{KZ_PS_AI}
  (\tau_Q^\text{KZ})_{\text{AI}}\sim\left.|g_f- g_c\right.|^{-1-z\nu}.
\end{equation}
as a turning point between the KZ and PS regimes.
Second, by setting $t_i=t_c-\hat{t}$, we get obtain another quench time scale,
\begin{equation}\label{PS_S_AI}
  (\tau_Q^\text{S})_{\text{AI}}\sim\left.|g_i-g_c\right.|^{-1-z\nu}.
\end{equation}
as a turning point between the PS and S regimes. The assumption in Eq. (\ref{assumption}) ensures the existence of the intermediate PS regime from the point of view of the AI approximation.

\section{Transverse Ising chain and quench protocol}\label{Subsec-Ising}

As a prototypical model of a quantum phase transition, we consider the transverse field quantum Ising chain,
\begin{align}\label{H-Ising}
H=-J \sum_{j=1}^{N}\left( \sigma_{j}^{x}\sigma_{j+1}^{x}+g\sigma_{j}^{z}\right),
\end{align}
where $\sigma^{a}_j$ ($a=x,y,z$) are Pauli matrices and the total number of lattice sites $N$ is assumed to be even. We impose a periodic boundary condition, $\sigma^{a}_{N+j}=\sigma^{a}_{j}$, and consider only the ferromagnetic case (i.e. $J>0$). We will set the reference energy scale to $J=1$ so that the strength of the transverse field is measured by $g$. By the Jordan-Wigner mapping, $\sigma_j^z=1-2c_j^{\dagger}c_j$ and $\sigma_j^x=-(c_j^{\dagger}+c_j)\prod_{l<j}\sigma_l^z$, and the canonical Bogoliubov transformation, $c_{q}=u_{q}\eta_{q}-v_{q}\eta_{-q}^{\dagger}$ with the Bogoliubov coefficients $u_{q}$ and $v_{q}$, we can arrive at the diagonalized form of the Hamiltonian in the quasiparticle representation,
\begin{align}\label{even Hamiltonian}
H=\sum_{q}\omega_{q}(\eta_{q}^{\dagger}\eta_{q}-\frac{1}{2}),
\end{align}
where $\eta_{q}$ is the quasiparticle operator, $q$ the quasimomentum, and $\omega_{q}=2\sqrt{1+g^2-2g \cos q}$ the quasiparticle dispersion.

In the thermodynamic limit $N\rightarrow\infty$ and at zero temperature, there is a second-order quantum phase transition from a ferromagnetic state ($0<g<1$) with $\mathbb{Z}_{2}$ symmetry breaking to a quantum paramagnetic state ($g>1$) \cite{Sachdev_2011}. The QCP occurs at $g_c = 1$, where the quasiparticle dispersion becomes a linear one, $\omega_q\sim 2|q-q_c|$ with critical quasimomentum $q_c=0$, that is responsible for the dynamical exponent $z=1$ and implies the correlation length exponent $\nu=1$.

We ramp linearly the transverse field from the paramagnetic to the ferromagnetic phases across the quantum critical point at a rate characterized by the quench time $\tau_Q$,
\begin{equation} \label{onewayg}
  g(t)=-\frac{t}{\tau_Q}, ~~(t_i \leq t\leq t_f).
\end{equation}
where $\tau_Q$ is the quench time, $t_i=-g_i\tau_Q$ the initial time, $t_f$ the final time,  and $g_i$ the initial transverse field. The system is initially in its ground state at a large initial value ($g_i\gg1$) to ensure the state located at paramagnetic phase deeply. Finally, $g(t)$ is ramped down to zero at $t=t_f=0$ and the system gets excited from its instantaneous ground state. At the final time, the Hamiltonian Eq. (\ref{even Hamiltonian}) reaches the classical Ising limit, thus the total number of defects (or kinks) can be measured by the operator,
\begin{eqnarray}
  \mathcal{N} &=& \frac{1}{2}\sum_{j=1}^{N}\left(1-\sigma_j^x\sigma_{j+1}^x\right), \label{kinknumber}
\end{eqnarray}
over the final state, which is in fact the number of excited quasiparticles \cite{Dziarmaga_2005}.

As time evolves, the quantum state $|\psi(t)\rangle$, which gets excited from the instantaneous ground state, should follow the time-dependent Bogoliubov transformation
\begin{align}
c_{q}=u_{q}(t)\tilde{\eta}_{q}+v^{*}_{-q}(t)\tilde{\eta}_{-q}^{\dagger},
\end{align}
where the quantum state has to be annihilated by the Bogoliubov fermions $\tilde{\eta}_q$ at every instant: $\tilde{\eta}_q|\psi(t)\rangle=0$. In Heisenberg picture, the fermion operator and Bogoliubov quasiparticle operator should satisfy  $i\frac{\mathrm{d}}{\mathrm{d}t}\tilde{\eta}_{q}=0$ and $i\frac{\mathrm{d}}{\mathrm{d}t}c_{q}=\left[c_{q}, H\right]$ \cite{Dziarmaga_2005, Das_2021_2}.

we can arrive at the dynamical version of the time-dependent Bogoliubov-de Gennes (TDBdG) equations,
\begin{align}\label{time-BDG}
i\frac{\mathrm{d}}{\mathrm{d}t}
\left[
\begin{array}{ccc}
u_{q}(t)\\v_{q}(t)
\end{array}
\right] =
\left[
\begin{array}{ccc}
\epsilon_{q}(t)&\Delta_{q}\\\Delta_{q}&-\epsilon_{q}(t)
\end{array}
\right]
\left[
\begin{array}{ccc}
u_{q}(t)\\v_{q}(t)
\end{array}
\right],
\end{align}
where $\epsilon_q(t)=2\{g(t)-\cos q\}$ and $\Delta_q=2\sin q$. It can be solved exactly by mapping to the Laudau-Zener (LZ) problem \cite{Dziarmaga_2005, Zener_1932}. We need to solve this problem for the linear ramp and calculate the density of defects through the excitation probability in the final state of the system.

\begin{table}[b]
	\renewcommand{\arraystretch}{1.6} 
	\centering
	\begin{tabular}{p{1.8cm}<{\centering}|p{1.7cm}<{\centering}p{2.2cm}<{\centering}p{1.7cm}<{\centering}}
		\hline
		\hline
		&KZ regime $(\tau_Q>1)$&PS regime $(g_i^{-2}<\tau_Q<1)$&S regime $(\tau_Q<g_i^{-2})$\\
		\hline
	    $t_i=-g_i\tau_Q$  &$\left.|z_i\right.|\rightarrow \infty$  &   $\left.|z_i\right.|\rightarrow \infty$  &   $\left.|z_i\right.|\rightarrow 0$\\
		\hline
		$t_f=0$     &$\left.|z_f\right.|\rightarrow \infty$   &   $\left.|z_f\right.|\rightarrow 0$    &   $\left.|z_f\right.|\rightarrow 0$\\
		\hline
	\end{tabular}
	\caption{Three types of regimes determined by the combination of limits of $|z_i|$ and $|z_f|$ defined in Eqs. (\ref{zi}) and (\ref{zf}) respectively. The asymptote of the parabolic cylinder function follows from Eqs. (\ref{asy-largez1}) or (\ref{asy-largez2}) when $|z|\rightarrow\infty$, but follows from Eq. (\ref{asy-smallz}) when $|z|\rightarrow 0$.}
\label{table-1}
\end{table}

And then, the LZ excitation probability is given by
\begin{align}\label{define-pq}
  p_q =& \langle\psi(0)|\eta^\dagger_q\eta_q|\psi(0)\rangle\nonumber\\
  =&\left|\cos\frac{q}{2}u_q(0)-\sin\frac{q}{2}v_q(0)\right|^2
\end{align}
at $t=0$, where $|\psi(t)\rangle$ is quantum state, and $u_q(t)$ and $v_q(t)$ are solutions of Eq. (\ref{time-BDG}). Generally, the kink density is related with the average excitation probability
\begin{equation}\label{define-n}
  n =\lim_{N\rightarrow\infty}\frac{1}{N} \langle\psi(0)|\mathcal{N}|\psi(0)\rangle=\frac{1}{\pi}\int_{q>0}dq~p_q.
\end{equation}

In Appendix \ref{appendix-TDBdG}, the solutions in Eq. (\ref{time-BDG}) are expressed in terms of complex parabolic cylinder functions with a variable $z=2\sqrt{\tau_{Q}}\left(\frac{t}{\tau_{Q}}+\cos q\right)e^{i\pi/4}$. The variable $z$ varies with $t$ from $z_i$ to $z_f$ that are formulated as
\begin{equation}\label{zi}
  z_i\equiv z|_{t=t_i}=2\sqrt{\tau_{Q}}\left(-g_i+\cos q\right)e^{i\pi/4}
\end{equation}
and
\begin{equation}\label{zf}
  z_f\equiv z|_{t=t_f}=2\sqrt{\tau_{Q}}\cos q e^{i\pi/4}
\end{equation}
respectively.

\section{Quench dynamics}\label{Subsec-density}

By applying the asymptotes of the parabolic cylinder functions that are given in Eqs. (\ref{asy-largez1})-(\ref{asy-smallz}), we find the quench dynamics falls into one of the three regimes that are listed in Table \ref{table-1}. In the following, we show the behaviors of the density of defects in the three regimes.

\subsection{Kibble-Zurek Regime} \label{subsubSec-KZ}

In the KZ regime, characterized by the slow quench when $\tau_Q\gg1$, the well-known KZM accurately predicts the behavior of defect density. In this regime, the long-wave approximation is valid since only long-wave modes within the small interval of $q\lesssim\frac{1}{\sqrt{\pi\tau_Q}}\ll\frac{\pi}{2}$ contribute, while short wave modes are rarely excited when the system is driven across the critical point. Meanwhile, we have $|z_i|\gg 1$ and $|z_f|\gg 1$. According to the asymptotes guided in Table \ref{table-1}, the time-dependent Bogoliubov coefficients at $t=0$ are worked out as
\begin{align}\label{KZ-uvq}
  &|u_q(0)|^2= e^{-2\pi\tau_Q q^2},\\
  &u_q(0)v_q(0)^*= e^{-\pi\tau_Q q^2}\sqrt{1-e^{-2\pi\tau_Q q^2}}e^{i\phi_q}, \label{KZ-uqvqphi}
\end{align}
where the dynamical phase reads
\begin{align}\label{KZ-DPhase}
  \phi_q=\frac{\pi}{4}+2\tau_{Q}+q^{2}\tau_{Q}\left(\ln4\tau_{Q}+\gamma_{E}-2\right),
\end{align}
and $\gamma_E$ is the Euler gamma constant. There are two length scales in the KZ regime. The first length scale is the correlation length (also known as KZ length)
\begin{align}\label{KZ-length}
  \hat{\xi}_{\text{KZ}}=\sqrt{\tau_Q}
\end{align}
contained in $|u_q(0)|^2$ or $|v_q(0)|^2$. The second length scale is the one $\propto\sqrt{\tau_{Q}\left(\ln4\tau_{Q}+\gamma_{E}-2\right)}$ implied in the dynamical phase $\phi_q$. Observably, the second length scale is much longer than the KZ correlation length in the large $\tau_Q$ limit, but it vanishes as $\tau_Q$ approaches the boarder to the PS regime, $\tau_Q\rightarrow 1$. It is well-known that KZM determines the spectrum of excitations $p_q$ after the system crosses the critical point, and subsequent dephasing of the excited quasiparticle modes manifests through the dynamical phase $\phi_q$ \cite{Dziarmaga_2021}. Therefore, there is a significant difference in the dephasing process between $\tau_Q\rightarrow 1$ and $\tau_Q\rightarrow\infty$, which will be further discussed in Sec. \ref{Sec-coherent}.

In this regime, the spectrum of excitations features a Gaussian decay in quasimomentum, $p_q=e^{-2\pi\tau_Q q^2}$. Thus the density of defects is given by
\begin{equation}\label{kink-KZ}
  n=\frac{1}{2\pi\sqrt{2\tau_Q}},
\end{equation}
which decays as the inverse square root of $\tau_Q$.

\begin{figure}[!htbp]
  \begin{center}
		\includegraphics[width=2.95 in,angle=0]{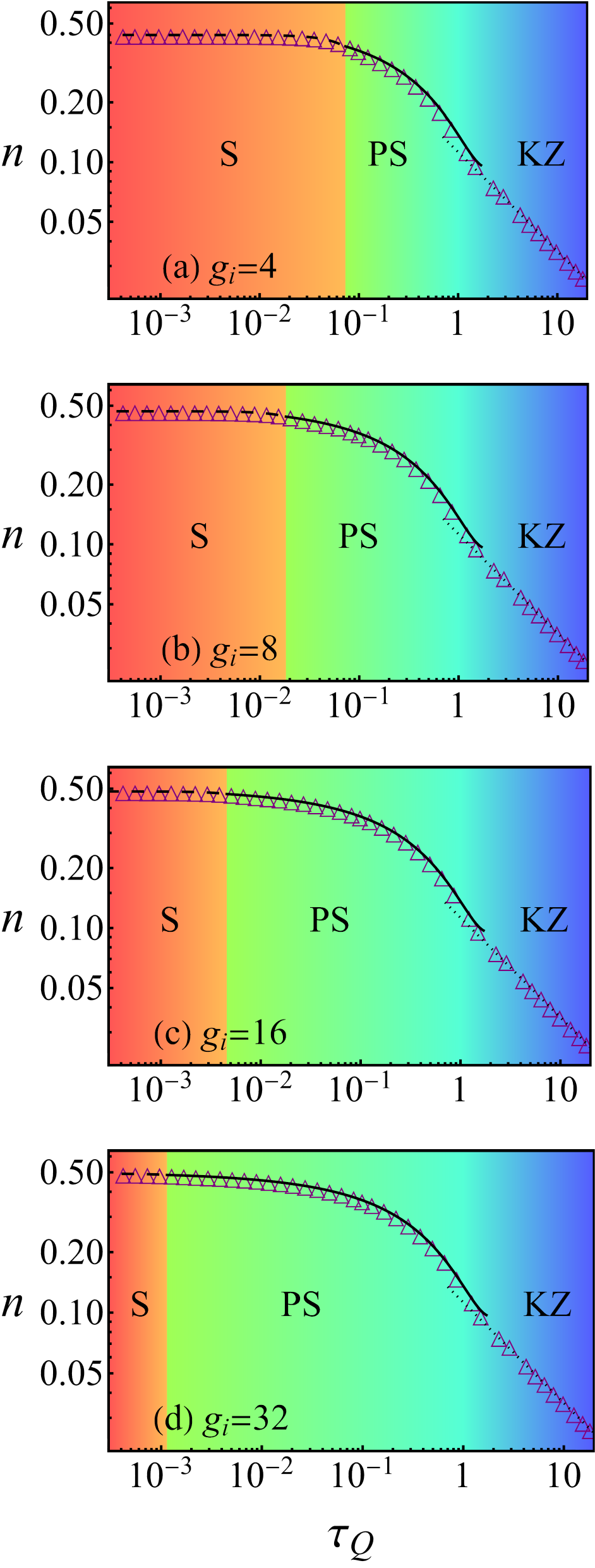}
  \end{center}
  \caption{The density of defects versus the quench time and the three regimes (S, PS, and KZ ones) with several selected parameters, $g_i=4$ (a), $8$ (b), $16$ (c) and $32$ (d). The dashed, solid, and dotted lines are generated by the analytical formulae in Eqs. (\ref{kink-S}), (\ref{kink-PS}), and (\ref{kink-KZ}) respectively, which are in very good agreement with the numerical solutions by Eq. (\ref{time-BDG}) denoted by the triangles. We can observe that the S regime shrinks rapidly with $g_i$ increasing and disappears eventually if $g_i\rightarrow\infty$.}
	\label{plot-kink}
\end{figure}

\subsection{Saturated regime}

If the evolution lasts only for a short period of time, breakdown of the KZ power law can be anticipated, which leads to a plateau in the defect density, $n_{\text{su}}+O(\tau_Q^{2})$, where $n_\text{su}$ is a constant attributed to a sudden quench \cite{Sun_2021}. There is a universality in the deviation from KZM \cite{Campo_2023}. In the S regime, characterized by a very fast quench with the condition, $\tau_Q< g_i^{-2}$, for a moderate or large initial transverse field, both $|z_i|$ and $|z_f|$ approach $0$. Following the prescription in Table \ref{table-1}, we work out two time-dependent Bogoliubov coefficients
\begin{align}
  &|u_{q}(0)|^2 = u_q(t_i)^2-\frac{4}{3}g_i^2\tau_Q^2\sin^2q,
\end{align}
\begin{align}
  u_q(0)v_q(0)^*=&u_q(t_i)v_q(t_i)+\frac{1}{3}g_i^3\tau_Q^2\sin q \nonumber\\
  &+ i(g_i+\cos q)\tau_Q\sin q
\end{align}
and the excitation probability
\begin{align}
  p_q= &p_q^\text{su}-\frac{1}{3}g_i^3\tau_Q^2\sin^2q\left\{1 +O(g_i^{-1})\right\}
\end{align}
where $u_q(t_i)$ and $v_q(t_i)$ are initial Bogoliubov coefficients formulated in Eqs. (\ref{initial-uq}) and (\ref{initial-vq}), and
\begin{equation}
p_q^\text{su}=\left|u_q(t_i)\cos\frac{q}{2}-v_q(t_i)\sin\frac{q}{2}\right|^2
\end{equation}
is the excitation probability in the sudden quench limit. Then we can get the final density of defects
\begin{align}
  &n=n_\text{su}-\frac{1}{6}g_i^3\tau_Q^2, \label{kink-S}\\
   n_\text{su}&=\left.\int_{0}^{\pi}\frac{dq}{\pi}p_q^\text{su}\right|_{g_i\gg1}=\frac{1}{2}-\frac{1}{4g_i}. \label{nsu}
\end{align}
The constant term, $\frac{1}{2}$, could be attributed to a sudden quench ($\tau_Q=0$) in the limit $g_i\rightarrow\infty$. The third term is a higher-order correction than the second one, since we have $g_i^3\tau_Q^2< g_i^{-1}$.

\subsection{Pre-saturated regime}

Now we consider another important situation. Herein, although the quench is fast ($\tau_Q<1$), but not so fast to exceed the square of the initial transverse field $g_i$ and we have $\tau_Q > g_i^{-2}$ instead of $\tau_Q < g_i^{-2}$. One can ensure this situation by preparing the initial system far from the critical point. In this case, we may consider the limits $|z_i|\rightarrow\infty$ and $|z_f|\rightarrow 0$ to search for appropriate asymptotes of the parabolic cylinder functions as prescribed in Table \ref{table-1} such that the two time-dependent Bogoliubov coefficients are worked out as
\begin{align}
  &|u_{q}(0)|^2 = \frac{|C_1|^2}{x} \left\{ 4~x^2 \sinh\frac{\pi x }{2}\cot^2q+\cosh\frac{ \pi x }{2} \right.\nonumber\\
  &~~~~~~~~~~~~~\left.-2~x \sqrt{\sinh \left(\pi x\right)} \cot q\right\},
\end{align}
\begin{align}
  &\label{PS-Re}
  u_q(0)v_q(0)^*= \frac{|C_1|^2}{\sqrt{x}} \left\{\sqrt{\frac{\sinh \left(\pi x\right)}{2 x}} e^{i\frac{\pi}{4}}\right.\nonumber\\
  &~~~~~~~~~~~~~\left. +\frac{(2x)^{\frac{3}{2}}}{\tan^2q}\sqrt{\sinh \left(\pi  x\right)} e^{i\frac{\pi}{4}}+i\frac{2 x e^{-\frac{\pi x}{2}}}{\tan q}\right\},
\end{align}
where $x=\tau_Q \sin^2q$ and $|C_1|^2$ is to be found in Eq. (\ref{C1-KZ}). To get an analytical result, the excitation probability defined in Eq. (\ref{define-pq}) is expanded into powers of $\tau_Q$ and, by keeping the lowest order $\sqrt{\tau_Q}$, we arrive at
\begin{align}\label{pq-PS}
  p_q=& \frac{1}{2} -\frac{1}{2} \cos q+u_q(t_i)^2 \cos q \nonumber\\
      &-\frac{\sqrt{\pi}}{2}\sqrt{\tau_Q} u_q^2(t_i) \sin ^2 q.
\end{align}

In contrast to the Gaussian decay observed in the KZ regime, the excitation probability exhibits a slower decay behavior as $q$ increases. To the order of $\tau_Q^{3/2}$, it is readily to verify that the density of defects can be worked out as
\begin{equation}\label{kink-PS}
  n= \frac{1}{2}-A(g_i)~\tau_Q^{1/2}+B(g_i)~\tau_Q^{3/2},
\end{equation}
where
\begin{align}
   & A(g_i)=\left(1-\frac{3}{16 g_i^2}\right)\frac{\sqrt{\pi}}{4} \label{Agi}\\
   & B(g_i)=\frac{\sqrt{\pi }}{32 g_i^2}-\frac{5 \pi ^{3/2}}{256 g_i^2}-\frac{\sqrt{\pi }}{4}+\frac{3 \pi ^{3/2}}{32}.\label{Bgi}
\end{align}
As the common feature with the S regime, the constant term, $\frac{1}{2}$, also originates from the sudden quench from a fully polarized paramagnetic state to a classical ferromagnetic state in the limit $g_i\rightarrow\infty$, although now we demand the condition, $\tau_Q > g_i^{-2}$.

To view a panorama of the S, PS, and KZ regimes from the slow to fast quench limits, we solve Eq. (\ref{time-BDG}) numerically and compare the numerical result with the above analytical ones at several selected transverse field $g_i=4$, $8$, $16$ and $32$. The comparison is illustrated in Fig. \ref{plot-kink}. Besides the KZ regime, the predictions in Eqs. (\ref{kink-S}) and (\ref{kink-PS}) are in very good agreement numerical solution in the S and PS regimes.

\subsection{Turning points}\label{Subsec-tpoint}

\begin{figure}[t]
  \begin{center}
	\includegraphics[width=3.3in,angle=0]{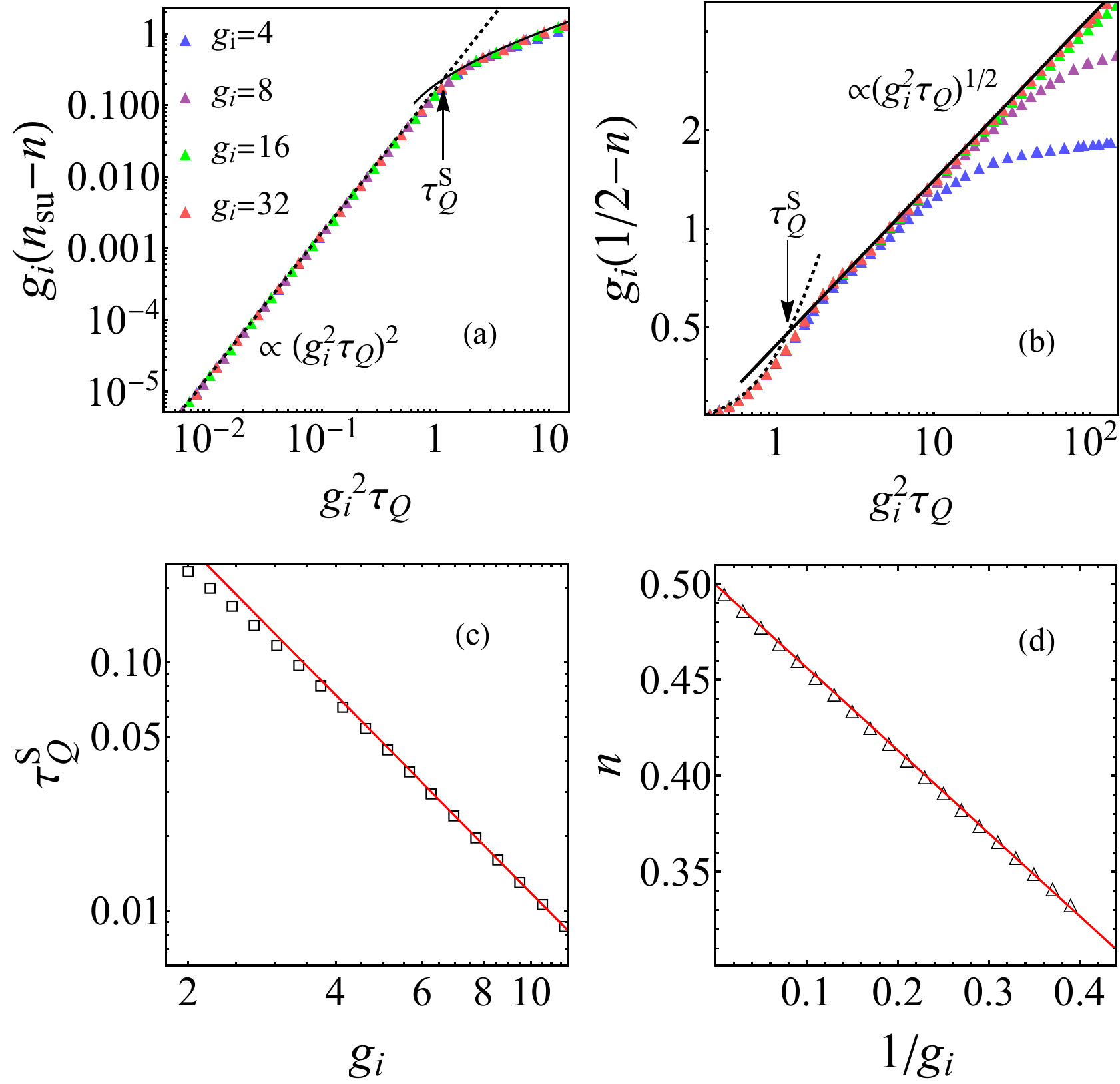}
  \end{center}
  \caption{Scaling behavior of defect density near the turning point from S to PS regimes: (a) $(n_\text{su}-n)g_i$ versus $g_i^2 \tau_Q$ according to Eq. (\ref{kink-S}) and (b) $(1/2-n)g_i$ versus $g_i^2 \tau_Q$ according to Eq. (\ref{kink-PS}). In (a), the numerical data collapse to the scaling law (dashed line), $g_i(n_\text{su}-n)\propto(g_i^2 \tau_Q)^2$. While in (b), the numerical data collapse to the scaling law (solid line), $g_i(1/2-n)\propto(g_i^2 \tau_Q)^{1/2}$, near the turning point. (c) The turning point $\tau_Q^{\text{S}}$ versus $g_i$ showing a power-law scaling behavior, $\tau_Q^{\text{S}}=1.17 g_i^{-2}$.  (d) Density of defects versus $1/g_i$ at the turning point $\tau_Q^{\text{S}}$, which reaches the value $1/2$ when $g_i\rightarrow\infty$. By fitting the data, we obtain  $n|_{\tau_Q^\text{S}}=1/2-0.43/g_i$.}
  \label{plot-root}
\end{figure}

From the above results, we see that the quench dynamics of the one-dimensional transverse Ising chain falls into one of three distinct regimes from the slow to fast limits. Now we look for the turning points between the regimes.

\subsubsection{Turning between S and PS regimes}

As outlined in Table. \ref{table-1}, there is a turning point between the S and PS regimes, whose defect densities are described by. Eqs. (\ref{kink-S}) and (\ref{kink-PS}) respectively. Near the turning point, the term with order $\tau_Q^{3/2}$ in Eq. (\ref{kink-PS}) can be neglected compared with the term with order $\tau_Q^{1/2}$ since we have $A(g_i)/B(g_i)= 5.61+O(g_i^{-2})$ according to Eqs. (\ref{Agi}) and (\ref{Bgi}). As illustrated in Fig. \ref{plot-root} (a) and (b), one can observe an obvious change of scaling behavior near a turning point $\tau_Q^S$. We can take the intersection point of the two curves in Eqs. (\ref{kink-S}) and (\ref{kink-PS}) as the turning point, which is obviously dependent on $g_i$. Then, from Fig. \ref{plot-root}(c), one can observe and verify numerically a scaling law,
\begin{equation} \label{tauS}
  \tau_Q^\text{S}=1.17 g_i^{-2},
\end{equation}
for large enough $g_i$. Consistently, according to Eq. (\ref{assumption}), the existence of PS regime is ensured by $g_i>2$ since we have $g_c=1$ and $g_f=0$ here. In Fig. \ref{plot-root}(d), we show the behavior of the defect density at $\tau_Q^{\text{S}}$ as a function of $1/g_i$. And by fitting the data, we obtain $n|_{\tau_Q^\text{S}}=1/2-0.43g_i^{-1}$. More interesting, the scaling behavior of the turning point implies that the S regime shrinks with $g_i$ increasing until it disappears in the limit $g_i\rightarrow \infty$ so that the PS regime dominates the entire fast quench regime.

\subsubsection{Turning between PS and KZ regimes}\label{subsec-secondpoint}

According to Eq. (\ref{kink-PS}), the defect density in the PS regime loses scaling behavior near the boarder to KZ regime since its third term proportional to $\tau_Q^{3/2}$ becomes significant. Meanwhile, according to Eq. (\ref{pq-PS}), the long-wave approximation fails because the modes with large quasimomentum $q$ are involved.

On the other hand, the dephasing effect in the KZ regime has an impact on the kink-kink correlation function through a dephasing length,
\begin{align}\label{dephasinglength}
  &l= \hat{\xi}_\text{KZ}\sqrt{1+\left[\frac{3}{4\pi}\left(\ln4\tau_{Q}+\gamma_{E}-2\right)\right]^2},
\end{align}
that describes the kink-kink correlation range \cite{Dziarmaga_2021}. The dephasing effect is a consequence of the interplay between the correlation length $\hat{\xi}_\text{KZ}$ and the second length in the dynamical phase expressed in Eq. (\ref{KZ-DPhase}). In the KZ regime, $l$ is much longer than the correlation length $\hat{\xi}_\text{KZ}$ for slow quench. But near the PS regime, it decreases and becomes comparable with the correlation length $\hat{\xi}_\text{KZ}$ and the dephasing effect is negligible. So, according to Eq. (\ref{KZ-DPhase}) and (\ref{dephasinglength}), we can take the value
\begin{equation}\label{tauQKZ}
  \tau_Q^\text{KZ}=\frac{e^{2-\gamma_{E}}}{4}\approx 1.037
\end{equation}
as the turning point between PS and KZ regimes (Please see Fig. \ref{plot-kink}). We notice it does not depend on the initial transverse field $g_i$. At this point, the dynamical phases of the different excited modes become independent of the quasimomentum $q$ since we have $\phi_q|_{\tau_Q^\text{KZ}}=\pi/4+2\tau_Q^\text{KZ}$. Moreover, a novel decay behavior in the kink-kink correlation is induced when entering into the PS regime, which will be demonstrated in the next section.

By Eqs. (\ref{KZ_PS_AI}) and (\ref{PS_S_AI}), we get the estimations, $(\tau_Q^\text{KZ})_{\text{AI}}\sim 1$ and $(\tau_Q^\text{S})_{\text{AI}}\sim g_i^{-2}$, which are in good agreement with the results in Eqs. (\ref{tauS}) and (\ref{tauQKZ}) respectively.

\section{Kink-Kink Correlation}\label{Sec-kkcor}

\begin{figure}[t]
  \begin{center}
		\includegraphics[width=3.1in,angle=0]{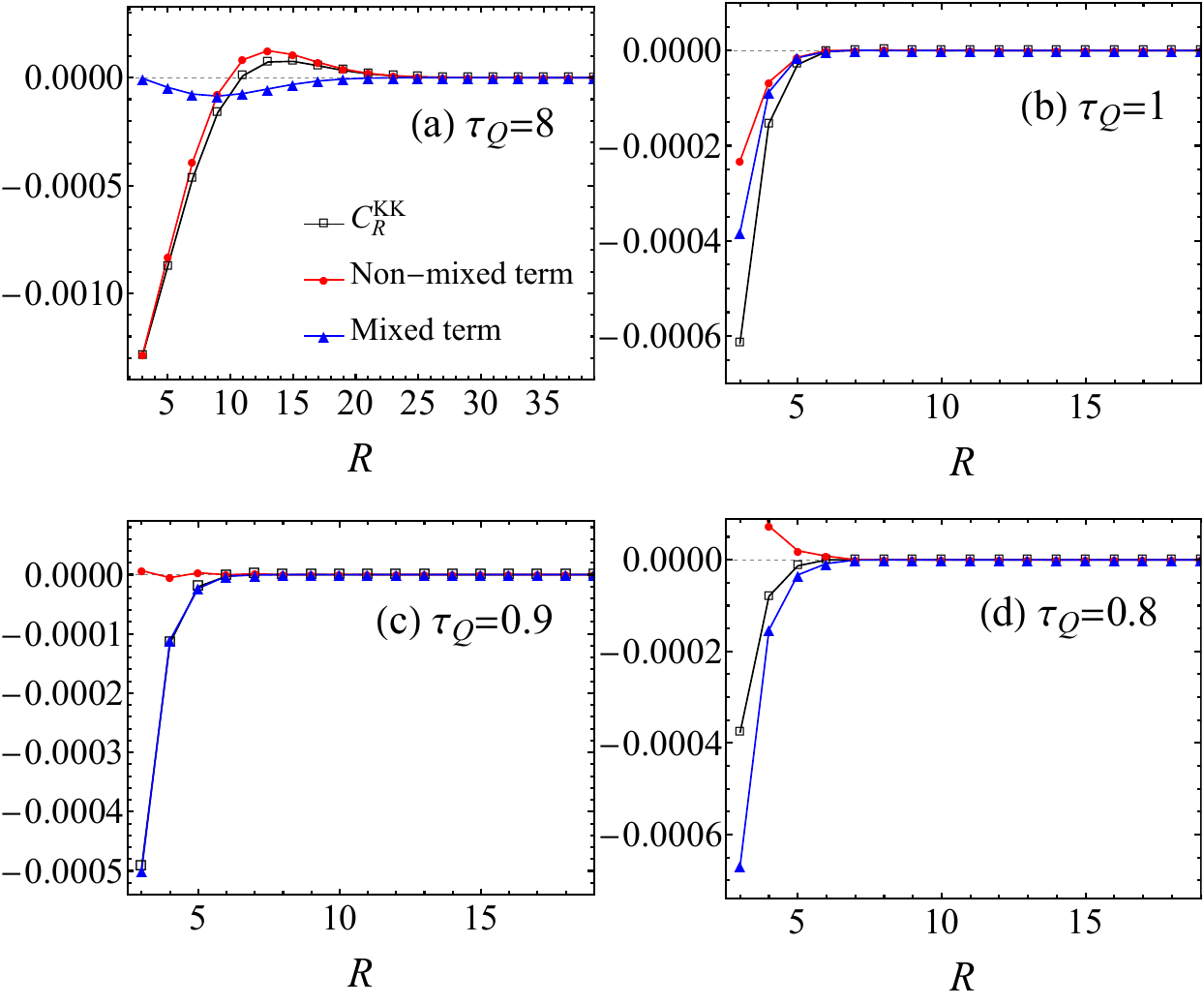}
  \end{center}
  \caption{Kink-kink correlator, non-mixed term and mixed term for various quench time $\tau_Q=8$~(a), $1$~(b), $0.9$~(c) and $0.8$~ (d), with $g_i=6$. In the KZ regime, the kink-kink correlator is mainly determined by the non-mixed term. However, we observe that the mixed term dominates the behavior of the kink-kink correlator in the PS regime. }
	\label{plot-correlator}
\end{figure}

In this section, we discuss the two-point correlation function between two defects. At $t=0$, the connected kink-kink correlation function between two kinks with distance $R$ is defined as
\begin{equation}
  C_R^{\text{KK}}=\left\langle K_j K_{j+R}\right\rangle-\left\langle K_j \right\rangle\left\langle K_{j+R}\right\rangle,
\end{equation}
where $K_j=\frac{1}{2}(1-\sigma_j^x\sigma_{j+1}^x)$ represents the kink number operator on the bond between sites $j$ and $j+1$. In the fermionic representation, the correlation can be expressed in terms of the diagonal and off-diagonal quadratic correlators and worked out as
\begin{align}
  \label{kink-corR}
  C_{R>1}^{\text{KK}} =& \text{Re}\beta_{R+1}\text{Re}\beta_{R-1}+(\text{Im}\beta_{R})^2-\alpha_{R+1}\alpha_{R-1}\nonumber\\
  &+\alpha_{R-1}\text{Re}\beta_{R+1}-\alpha_{R+1}\text{Re}\beta_{R-1},\\
  \label{kink-cor1}
  C_{R=1}^{\text{KK}} =& (\text{Im}\beta_{1})^2-\frac{\text{Re}\beta_2 }{2} -\alpha_{2}\alpha_{0}+\alpha_{0}\text{Re}\beta_{2}+\frac{\alpha_2}{2},\\
  \label{kink-cor0}
  C_{R=0}^{\text{KK}} =& \frac{1}{4}-(\text{Re}\beta_{1})^2-\alpha_{1}^2+2\alpha_{1}\text{Re}\beta_{1},
\end{align}
where \begin{align}\label{alphaR}
  \alpha_R =&\frac{2}{N}\sum_{q>0}\left.|u_q(t)\right.|^2\cos(qR), \\
  \label{betaR}
  \beta_R =&\frac{2}{N}\sum_{q>0}u_q(t)v_q^*(t)\sin(qR),
\end{align}
are the diagonal and off-diagonal correlators respectively. In Eq. (\ref{kink-corR}), we can discern the non-mixed terms and mixed terms. The non-mixed terms include the off-diagonal ones, $\text{Re}\beta_{R+1}\text{Re}\beta_{R-1}$ and $(\text{Im}\beta_{R})^2$, that only contain off-diagonal correlators and the diagonal one, $-\alpha_{R+1}\alpha_{R-1}$, that contains only diagonal correlator. The mixed terms, $\alpha_{R-1}\text{Re}\beta_{R+1}$ and $-\alpha_{R+1}\text{Re}\beta_{R-1}$, contain both diagonal and off-diagonal correlators. In Fig. \ref{plot-correlator}, we exhibit the contributions of the terms. In the KZ regime, one can observe that the mixed terms can be neglected\cite{Dziarmaga_2021} (Fig. \ref{plot-correlator}(a)). However, after entering into the PS regime, the mixed terms take in charge (Fig. \ref{plot-correlator}(b)-(d)).

\begin{figure}[t]
  \begin{center}
		\includegraphics[width=3.18in,angle=0]{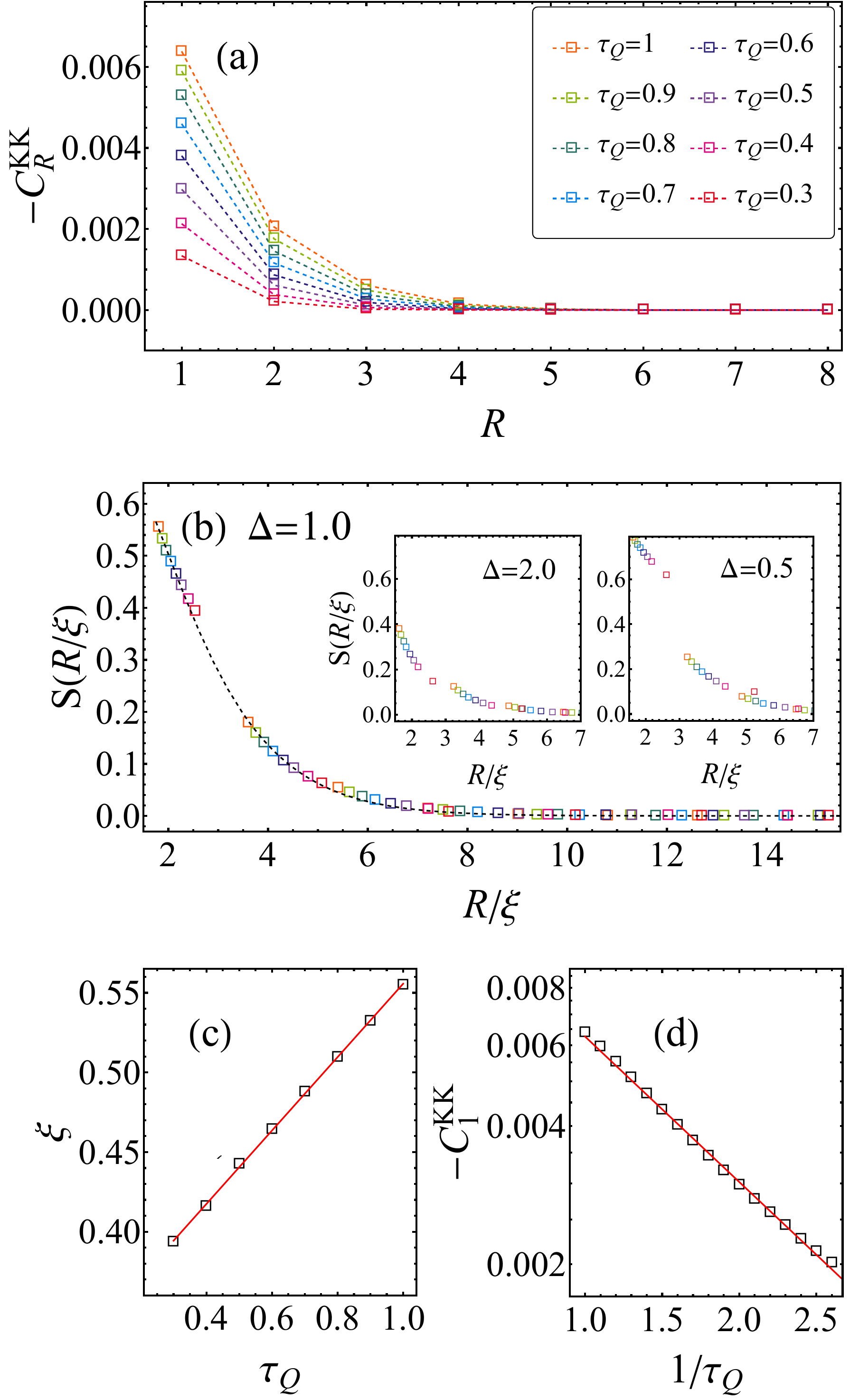}
  \end{center}
  \caption{(a) Kink-kink correlation versus the distance $R$, for several values of the quench time, $\tau_Q$. We select the parameter $g_i=8$ here. (b) Non-universal scaling function $S(R/\xi)$ versus the scaled distance $R/\xi$. The numerical data collapse to the conjectured scaling function in Eq. (\ref{S-function}) (denoted by the dashed line). The inset shows the scaling function can not fit for the data uniformly if $\Delta$ deviates from 1. (c) The correlation length. (d) The kink-kink correlation at $R=1$. In (c) and (d), we obtain $\xi=0.33+0.23\tau_Q$ and $C_1^{\text{KK}}=-0.013 e^{-0.73/\tau_Q}$, by fitting the data.}
	\label{plot-KKCscaling}
\end{figure}

As a consequence, the behavior of the kink-kink correlation undergoes a significant change during the breakdown of the KZ scaling law from KZ to PS regimes. In the KZ regime, the kink-kink correlation features a Gaussian decay \cite{Das_2021_2, Dziarmaga_2021, Kou_2022},
\begin{equation}
  n^{-2}C_R^{\text{KK}}= a\frac{\hat{\xi}_\text{KZ}R^2}{l^3}e^{-3\pi(R/l)^2}-e^{-2\pi(R/\hat{\xi}_\text{KZ})^2},
\end{equation}
where $n$ is formulated as Eq. (\ref{kink-KZ}), and $a=9.75$ is a numerical prefactor.  However, near the boarder to the PS regime, the length $l$ shrinks to a scale comparable with the correlation length $\hat{\xi}_\text{KZ}$, so the kink-kink correlation is expected to deviate from the Gaussian decay. While in the PS regime, we adopt a scaling hypothesis,
\begin{equation}\label{KKC-scaling}
  C_R^{\text{KK}}/ C_1^{\text{KK}}= \xi^{-\Delta}S(R/\xi),
\end{equation}
where the unknown $\xi$ is the correlation length for this regime, $\Delta$ is the scaling dimension and $S(x)$ is a non-universal scaling function \cite{Dziarmaga_2023}. First, $C_1^{\text{KK}}$, defined in Eq. (\ref{kink-cor1}), is independent of the distance $R$ and can be fitted numerically alone. As outlined in Fig. \ref{plot-KKCscaling}(d), we find it is described by $C_1^{\text{KK}}\approx-0.013 e^{-0.73/\tau_Q}$ quite well. Second, the non-universal scaling function is conjectured tentatively as $(R/\xi)e^{-R/\xi}$ so that we can extract the correlation length $\xi$. By varying $\Delta$, it can be observed whether the data collapse to the conjectured scaling function. From Fig. \ref{plot-KKCscaling}(a) and (b), we see the data collapse to the scaling function quite well when $\Delta\approx1$, where we obtain
\begin{equation}\label{S-function}
  S(R/\xi)=1.86 e^{-R/\xi}R/\xi
\end{equation}
and
\begin{equation}
  \xi=0.33+0.23\tau_Q.
\end{equation}

\section{Coherent Many-Body Oscillation}\label{Sec-coherent}

Finally, we investigate the coherent many-body oscillation after the quench, which is complementary to the dephasing, as a detecting means to measure the dephasing effect on the superposition state \cite{Dziarmaga_2022}. After the system is quenched across the critical region, its post-transition state is a superposition of states that populates with topological defects. The superposition inevitably results in the quantum coherent oscillation. In the KZ regime, the coherent quantum oscillation satisfies a Kibble-Zurek dynamical scaling laws. It is interesting to explore the behavior of the coherent many-body oscillation in the PS and S regimes.

We calculate time-dependent transverse magnetization. It is given by the expression
\begin{equation}
  \langle \sigma_j^z(t)\rangle = \langle e^{i\int_{t}H(t')dt'}\sigma_j^z e^{-i\int_{t}H(t')dt'}\rangle_t,
\end{equation}
where the system freely evolves after a linear quench,
\begin{equation}\label{halt-g}
  g(t)=\left\{
  \begin{array}{cl}
    -\frac{t}{\tau_Q}, &~~ -g_i\tau_Q<t<0, \\
    0, &~~ t\geq0.
  \end{array}
  \right.
\end{equation}
For the free evolution $t>0$, the transverse magnetization can be worked out as
\begin{equation}
  \langle \sigma_j^z(t)\rangle=A+M\cos(4t+\phi),
\end{equation}
which exhibits a coherent oscillation with a period $T=\frac{\pi}{2}$ along the time $t$. The non-oscillatory part, the amplitude, and the phase angle respectively read
\begin{align}
  \label{sigmaz-A}
  &A=\alpha_0+\alpha_2-\text{Re}\beta_2-\frac{1}{2},\\
  \label{sigmaz-M}
  &M^2=\left(\alpha_0-\alpha_2+\text{Re}\beta_2-\frac{1}{2}\right)^2+4\left(\text{Im}\beta_1\right)^2,\\
  &\tan\phi=\frac{2\text{Im}\beta_1}{\alpha_0-\alpha_2+\text{Re}\beta_2-\frac{1}{2}},
\end{align}
where $\alpha_R$ and $\beta_R$ are defined in Eqs. (\ref{alphaR}) and (\ref{betaR}) at $t=0$.

\begin{figure}[t]
  \begin{center}
		\includegraphics[width=3.3in,angle=0]{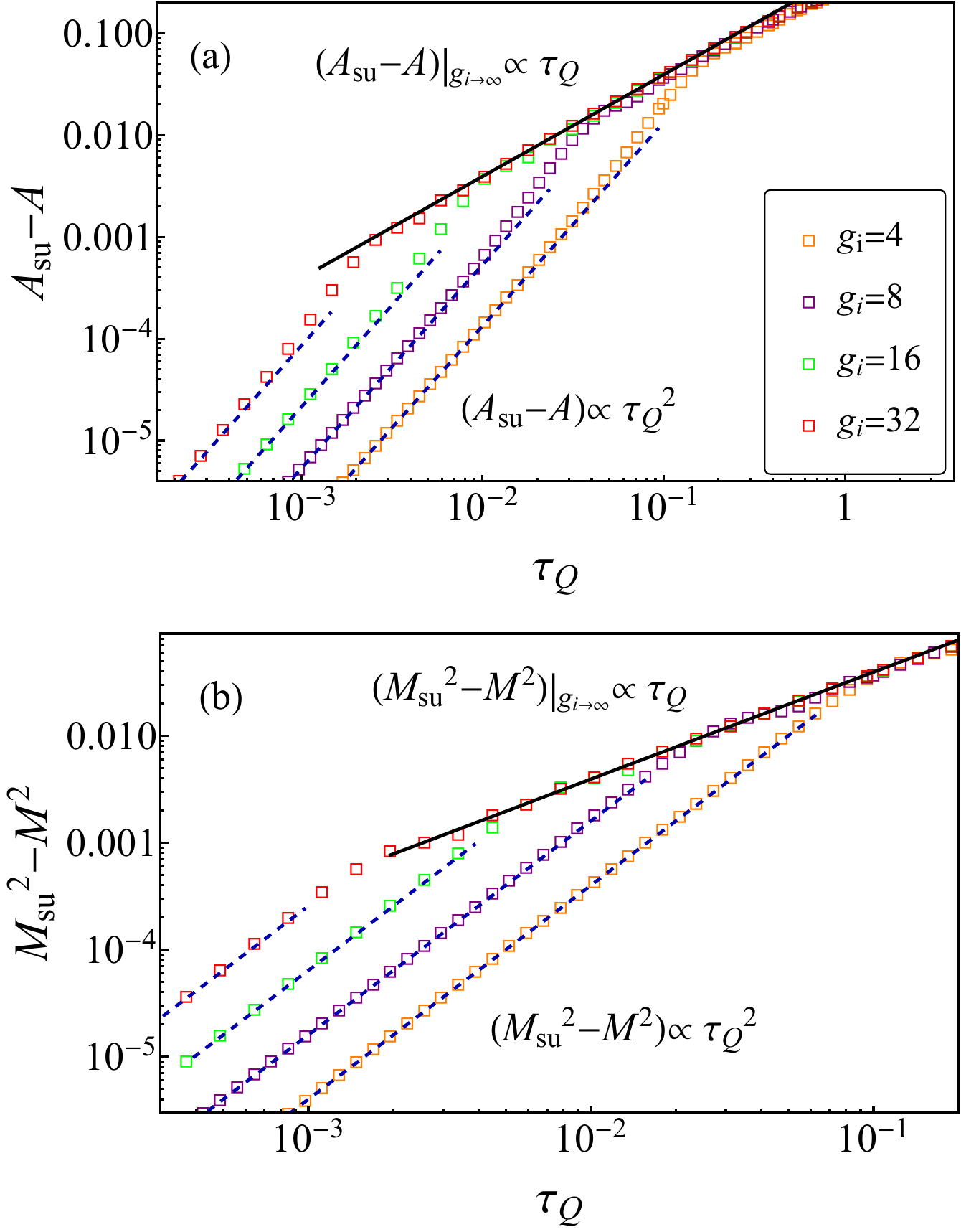}
  \end{center}
  \caption{The scaling behaviors of $A_\text{su}-A$ (a) and $M_\text{su}^2-M^2$ (b) as a function quench time with several selected parameters, $g_i=4$, $8$, $16$ and $32$. $A_\text{su}-A$ and $M_\text{su}^2-M^2$ scale as $\propto\tau_Q^2$ in the S regime, and, for sufficiently large $g_i$, they scale as $\propto\tau_Q$ in the PS regime. The lines are analytical results according to the formulae in Eqs. (\ref{sigmaz-A2}) and (\ref{sigmaz-Msquare2}). }
	\label{plot-AA}
\end{figure}

The final result for the KZ regime was given by Dziarmaga et al. in Ref. \cite{Dziarmaga_2022}. Here, we focus on the results for the S and PS regimes, which read
\begin{equation} \label{sigmaz-A2}
  A=\left\{
  \begin{array}{lc}
    \frac{1}{2}-\frac{1}{16 g_i^2}+\left(\frac{\pi }{64 g_i^2}-\frac{\pi }{8}\right) \tau_Q, & \text{PS regime}, \\
    \frac{1}{2}-\frac{3}{16 g_i^2}-\frac{g_i^2}{12}\tau_Q^2,   & \text{S regime},
  \end{array}
  \right.
\end{equation}
and
\begin{equation}\label{sigmaz-Msquare2}
  M^2=\left\{
  \begin{array}{lc}
    \frac{1}{4}-\frac{3}{16 g_i^2}+\left(\frac{7 \pi }{32 g_i^2}-\frac{\pi }{8}\right) \tau_Q, & \text{PS regime}, \\
    \frac{1}{4}-\frac{1}{16 g_i^2}-\frac{1}{4} g_i^2 \tau_Q^2,  & \text{S regime}.
  \end{array}
  \right.
\end{equation}
In the case of a sudden quench limit, i.e. $\tau_Q=0$, we have $A_\text{su}=\frac{1}{2}-\frac{3}{16 g_i^2}$ and $M^2_\text{su}=\frac{1}{4}-\frac{1}{16 g_i^2}$. The non-oscillatory part and amplitude scale as $\mathcal{O}_\text{su}-\mathcal{O}\propto\tau_Q^2$ where $\mathcal{O}$ represents either $A$ or $M^2$ in the S regime. However, in the large initial transverse field limit, $g_i\rightarrow\infty$, they scale as $\mathcal{O}_\text{su}-\mathcal{O}\propto\tau_Q$ in the PS regime. Thus, there is a change of scaling behaviors near the vicinity of $\tau_Q^\text{S}$. These analytical results are confirmed by numerical ones, as illustrated in Fig. \ref{plot-AA}.

\begin{figure}[t]
  \begin{center}
		\includegraphics[width=3.23in,angle=0]{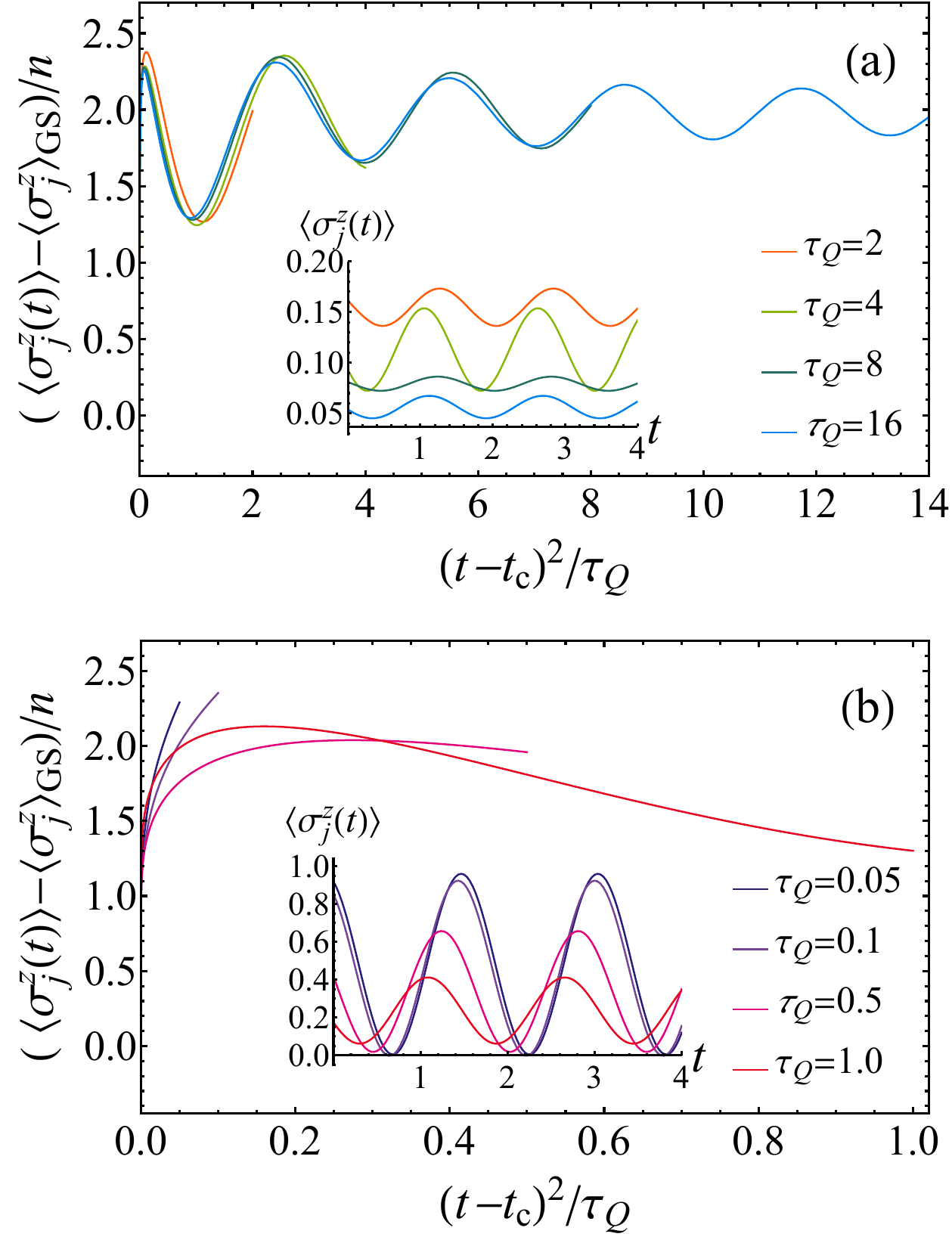}
  \end{center}
  \caption{The time-dependent magnetization with fixed parameter $g_i=6$ for the slow quench (a) and the fast quench (b). The scaled data is a function of scaled time, $(t-t_c)^2/\tau_Q$. In (a), the data collapse to the same curve and shows the coherent oscillation with fixed period given by Eq. (\ref{period-magnetization}), which causes the dephasing process. In (b), the data show that the evolution lacks dephasing process. The insets of (a) and (b) show that a longer duration of dephasing results in a weaker coherent oscillation during free evolution. }
	\label{plot-dephasing}
\end{figure}

Furthermore, in the kink-kink correlator, we have observed a shrink of the characteristic length $l$ from the KZ to PS regime, which is due to an attenuation of the dephasing effect. Here, we demonstrate that it can also be observed in the time-dependent magnetization. After the critical quench dynamics, the off-diagonal correlator exhibits a coherent oscillation that leads to the dephasing effect \cite{Dziarmaga_2010, Kou_2022},
\begin{equation}\label{time-betaR}
  \beta_R(t)=\frac{e^{\frac{2 i t^2}{\tau_Q}}}{2\pi i}\int_{-\pi}^{\pi}dq \left|u_q(t)v_q(t)^*\right|e^{-i(4t\cos q+2\phi_q-qR)},
\end{equation}
where $\phi_q$ is dynamical phase and is formulated as Eq. (\ref{KZ-DPhase}) for the  KZ regime. The dephasing effect is the interplay of these two terms. One is $e^{2 i t^2/\tau_Q}$ that causes a coherent oscillation with period $T_{t^2}=\pi\tau_Q$ along the axis of square of time. The other is $e^{-i4t\cos q}$ that results in a amplitude decay after integration. The depahsing effect also occurs in the time-dependent maganetization,
\begin{equation}\label{period-magnetization}
  \left(\langle \sigma_j^z(t)\rangle-\langle \sigma_j^z\rangle_\text{GS}\right)/n\sim \cos\left(\frac{2}{\tau_Q}t^2\right).
\end{equation}
where $\langle \sigma_j^z\rangle_\text{GS}$ is the transverse magnetization of the ground state. The total evolution time of linear quench is in proportion to the quench time (i.e., $t_f-t_i=g_i\tau_Q$), so a larger quench time leads to a longer dephasing time. In the interval, $t_c<t<0$ where $t_c=-\tau_Q$ is the time when the system crosses the critical point, the slow quench exhibits a longer duration of dephasing with oscillation in the fixed period $T_{t^2}$ as illustrated in Fig. \ref{plot-dephasing} (a). But, the fast quench exhibits a transient duration of dephasing, as illustrated in Fig. \ref{plot-dephasing} (b). As shown in the insets in the Fig. \ref{plot-dephasing} (a) and (b), when $t>0$, the amplitude $A$ gradually decreases as the duration of the dephasing increases. This means the dephasing effect weaken the coherence inevitably.

\section{Summary}\label{Sec-discussion}

In summary, we have demonstrated that there can be a PS regime lying between the S and KZ regimes. The scenario is established according to the AI approximation first. Then we provide both analytical and numerical calculations on the transverse field Ising model to realize the scenario. As we shift the quench dynamics from the S to PS regimes, the scaling behavior in the defect density changes from $(n-n_\text{su})g_i\propto(g_i^2\tau_Q)^{2}$ to $(n-1/2)g_i\propto(g_i^2\tau_Q)^{1/2}$ near turning point $\tau_Q^\text{S}$. This turning point scales with the initial transverse field, $\tau_Q^\text{S}\propto g_i^{-2}$, implying that the S regime vanishes for an infinite initial transverse field, $g_i\rightarrow\infty$. As we shift the quench dynamics from the KZ to PS regimes, the dephasing effect is attenuated. Near the turning point $\tau_Q^\text{KZ}$, the kink-kink correlator exhibits an exponential decay behavior rather than a Gaussian decay. This is due to the fact that the characteristic length $l$ shrinks to the scale of the KZ length $\hat{\xi}_\text{KZ}$. Below $\tau_Q^\text{KZ}$, the lacks of the dephasing effect leads to more prominent coherent oscillations in the post-quench state during free evolution.

Intriguingly, a significant PS regime can emerge after the KZ scaling law breaks down and before the S scaling law develops. Although this finding is mainly based on a prototypical integrable system, the scenario of the AI approximation suggests that the conclusion may be generalized to other systems, such as the non-integrable systems \cite{Sengupta_2012, Sengupta_2018}, to which the AI approximation is also applicable \cite{Zurek_2012, Zurek_2017, Yin_2020, Campo_2023}.

\section*{ACKNOWLEDGMENTS}

We thank Yan He for seminal discussion. It is also a pleasure to acknowledge discussions with G. Lamporesi, Y. Shin, Kyuhawan Lee, and Karin Sim. This work is supported by NSFC under Grants No. 11074177.

\appendix

\section{Solution of the  TDBdG equations} \label{appendix-TDBdG}

We can solve the TDBdG equations given by Eq. (\ref{time-BDG}) exactly by mapping them to the Landau-Zener problem. Then, the time-dependent Bogoliubov coefficients can be given by
\begin{align}
\label{solu-vq}
v_{q}(z)=&C_{1}D_{-s_{q}-1}(i z)+C_{2}D_{-s_{q}-1}(-i z),\\
\label{solu-uq}
u_{q}(z)=&\frac{e^{i\pi/4}}{\sqrt{\tau_{Q}}\sin q}\left(i\frac{\mathrm{d}}{\mathrm{d}z}+\frac{i z}{2}\right)v_{q}(z),
\end{align}
with free complex parameters $C_{1}$ and $C_{2}$. Here, $D_{m}(z)$ is the complex parabolic cylinder function, $z=2\sqrt{\tau_{Q}}\left(\frac{t}{\tau_{Q}}+\cos q\right)e^{i\pi/4}$, and $s_{q}=-i\tau_{Q}\sin^{2}q$. To reduce the above rigorous solution, we need to apply the asymptotes of $D_{m}(z)$ that are given by \cite{Frank_2010}
\begin{align}\label{asy-largez1}
	D_{m}(z)=e^{-z^{2}/4}z^{m},~\forall |\arg(z)|<3\pi/4,
\end{align}
\begin{align}\label{asy-largez2}
  D_{m}(z)=&e^{-z^{2}/4}z^{m}-\frac{\sqrt{2\pi}}{\Gamma(-m)}e^{-im\pi}e^{z^{2}/4}z^{-m-1},\nonumber \\
  &~\forall -5\pi/4<\arg(z)<-\pi/4,
\end{align}
for $|z|\gg 1$ and
\begin{align}\label{asy-smallz}
D_{m}(z)=\frac{2^{m/2}\sqrt{\pi}}{\Gamma(\frac{1}{2}-\frac{m}{2})}-\frac{2^{\frac{1}{2}+\frac{m}{2}}\sqrt{\pi}z}{\Gamma(-\frac{m}{2})}+O(z^2),
\end{align}
for $|z|\rightarrow 0$.

Furthermore, in numerical simulations, the time-dependent parameter should start at a finite value. We choose a sufficiently large but finite initial transverse field, so the initial conditions of Eqs. (\ref{solu-vq}) and (\ref{solu-uq}) can be expanded into a powers of $1/g_i$,
\begin{align}
  \label{initial-uq}
  u_q(t_i)^2&= 1-\frac{\sin^2q}{4g_i^2}+O(\frac{1}{g_i^3}), \\
  \label{initial-vq}
  v_q(t_i)^2&=1-u_q(t_i)^2.
\end{align}
Based on this approximation, the two constants, $C_1$ and $C_2$, can be expressed as
\begin{align}
	\label{C1-KZ}
    |C_{1}|^2&=u_q(t_i)^2~ e^{-\frac{\pi}{2}\tau_Q\sin^2q}\tau_{Q}\sin^2q, \\
    \label{C2-KZ}
    |C_{2}|^2&=0,
\end{align}
for $|z_i|\gg 1$, and
\begin{eqnarray}
  \label{C1-S}
  C_1 =& \frac{v_q(t_i)}{\sqrt{2\pi}}-\frac{(-1)^{3/4}u_q(t_i)\sqrt{\tau_Q}\sin q}{2}+O(\tau_Q,\tau_Q^2),\\
  \label{C2-S}
  C_2 = & \frac{v_q(t_i)}{\sqrt{2\pi}}+\frac{(-1)^{3/4}u_q(t_i)\sqrt{\tau_Q}\sin q}{2}+O(\tau_Q,\tau_Q^2),
\end{eqnarray}
for $|z_i|\ll 1$, where $z_i$ is defined by Eq. (\ref{zi}).

\bibliographystyle{aapmrev4-2}
\bibliography{citation}


\end{document}